\documentclass[aps,pra,twocolumn,groupedaddress,amssymb,amsmath,showpacs]{revtex4}
\usepackage{dsfont}
\usepackage{bm}
\usepackage{bbm}
\usepackage[mathscr]{euscript}
\usepackage{amsmath}
\usepackage{amssymb}
\usepackage{amsthm}
\usepackage{amsfonts}
\usepackage{graphicx}
\usepackage{color}
\usepackage[colorlinks=true,urlcolor=blue,linkcolor=blue,citecolor=red]{hyperref}

\begin{document}


\title{Entanglement Detection Using Majorization Uncertainty Bounds}


\author{M. Hossein \surname{Partovi}}
\email[Electronic address:\,\,]{hpartovi@csus.edu}
\affiliation{Department of Physics and Astronomy, California State
University, Sacramento, California 95819-6041}


\date{\today}

\begin{abstract}
Entanglement detection criteria are developed within the framework of the majorization formulation of uncertainty.   The primary results are two theorems asserting linear and nonlinear separability criteria based on majorization relations, the violation of which would imply entanglement.    Corollaries to these theorems yield infinite sets of scalar entanglement detection criteria based on quasi-entropic measures of disorder.  Examples are analyzed to probe the efficacy of the derived criteria in detecting the entanglement of bipartite Werner states.  Characteristics of the majorization relation as a comparator of disorder uniquely suited to information-theoretical applications are emphasized throughout.

\end{abstract}



\maketitle



\section{Introduction}
Quantum measurements in general have indeterminate outcomes, with the results commonly expressed as a vector of probabilities corresponding to the set of outcomes.  In case of noncommuting observables, the joint indeterminacy of their measurement outcomes has an inviolable lower bound, in stark contrast to classical expectations.   This was discovered by Heisenberg in his desire to advance the physical understanding of the newly discovered matrix mechanics by relating the unavoidable disturbances caused by the act of measurement to the fundamental commutation relations of quantum dynamics \cite{HEI}.  Heisenberg's arguments relied on the statistical spread of the measured values of the observables to quantify uncertainty.  This gave rise to the variance formulation of the uncertainty principle which remains a powerful source of intuition on the structure and spectral properties of microscopic systems.  With the prospect of quantum computing and the development of quantum information theory in recent decades, on the other hand, the need for a measure of uncertainty that can better capture its information theoretical aspects, especially in dealing with noncanonical observables, has inspired new formulations.  Among these are the entropic measure developed in the eighties, and the majorization formulation proposed recently.  Uncertainty relations resulting from these formulations have found application to quantum cryptography, information locking, and entanglement detection, in addition to providing uncertainty limits \cite{SUR,I}.

In this paper we develop applications of the majorization formulation of uncertainty introduced in Ref. \cite{I} to the problem of entanglement detection.  Deciding whether a given quantum state is entangled is a central problem of quantum information theory and known to be computationally intractable in general \cite{GUR}.  As a result, computationally tractable necessary conditions for separability, which provide a partial solution to this problem, have been the subject of active research in recent years.  Among these, the Peres-Horodecki positive partial transpose criterion actually provides necessary and sufficient separability conditions for $2\otimes2$ and $2\otimes3$ dimensional systems, and necessary conditions otherwise.  Other notable results are the reduction and global versus local disorder criteria, both necessary conditions in general \cite{sep}.  An observable that has non-negative expectation values for all separable states and negative ones for a subset of entangled states provides an operational method of entanglement detection and is known as an entanglement witness \cite{HOR}.   It has also long been known that uncertainty relations can serve a similar purpose by providing inequalities that must be satisfied by separable states and if violated signal entanglement \cite{GIO,GAM}.

In this paper we extend the majorization formulation of uncertainty developed in Ref. \cite{I} to the problem of entanglement detection.  As discussed in that paper (hereafter referred to as Paper I) and in the following, majorization as a comparator of uncertainty is qualitatively different from, and stronger than, scalar measures of uncertainty.  Consequently, the entanglement detection results developed here represent a qualitative strengthening of the existing variance and entropic results.  This will be evident, among other things, by the fact that they yield, as corollaries,  an infinite class of entanglement detectors based on scalar measures.

As mentioned above, results of quantum measurements are in general probability vectors deduced from counter statistics, and an information theoretical formulation of uncertainty is normally based on an order of uncertainty defined on such vectors (see Paper I).  Thus a scalar measure of uncertainty is commonly a real, non-negative function defined on probability vectors whose value serves to define the uncertainty in question.  For example, the Shannon entropy function is the measure of uncertainty for the standard entropic formulation of uncertainty \cite{DEU}.  By contrast, majorization provides a partial order of uncertainty on probability vectors that is in general more stringent, and fundamentally stronger, than a scalar measure \cite{maj}.  Note that, unlike scalar measures, majorization does not assign a quantitative measure of uncertainty to probability vectors, and as a partial order may find a pair of vectors to be incomparable.

A characterization of majorization that clarifies the foregoing statements can be attained by considering the quasi-entropic set of measures.  These were defined in Paper I as the set of concave, symmetric functions defined on probability vectors, and include the Shannon, Tsallis, and (a subfamily of) R\'{e}nyi entropies as special cases \cite{ftn0}.   It is important to realize that the uncertainty order determined by one member of the quasi-entropic set for a given pair of probability vectors may contradict that given by another.   While additional considerations may justify the use of, e.g., Shannon entropy in preference to the others, the foregoing observation clearly indicates the relative nature of the uncertainty order given by a specific measure, and immediately raises the following question: are there pairs of vectors for which all quasi-entropic measures determine the same uncertainty order?   The answer is yes, and the common determination of the quasi-entropic set in such cases defines the uncertainty order given by majorization.   What about the cases where there are conflicting determinations by the members of the quasi-entropic set?  The majorization relation defines such pairs as incomparable, whence the ``partial'' nature of the order defined by majorization.  We may therefore consider the majorization order to be equivalent to the collective order determination of the entire quasi-entropic set (see \S IV).   This characterization of the majorization relation clearly shows its standing vis-\`{a}-vis the scalar measures of uncertainty.  More practical definitions of the majorization relation will be considered in \S II.

The rest of this paper is organized as follows.  In \S II we review elements of the majorization formulation of uncertainty needed for our work and establish the notation.  In \S III we present the central results of this paper on entanglement detection, and in \S IV we derive entire classes of entanglement detectors for quasi-entropic measures as corollaries to the theorems of \S III.  Concluding remarks are presented in \S V.   Details of certain mathematical proofs are given in the Appendix.

\section{Majorization formulation of uncertainty}
This section presents a review of the basic elements of majorization theory and the formulation of the uncertainty principle based on it.  It follows the treatment given in Paper I.

\subsection{Majorization}
The basic element of our formulation is the majorization protocol for comparing the degree of uncertainty, or disorder, among probability vectors, i.e., sequences of non-negative numbers summing to unity \cite{maj}.   By definition, ${\lambda}^{1}$ is no less uncertain than ${\lambda}^{2}$ if ${\lambda}^{1}$ equals a mixture of the permutations of ${\lambda}^{2}$.  Then ${\lambda}^{1}$ is said to be majorized by ${\lambda}^{2}$ and written ${\lambda}^{1}\prec{\lambda}^{2}$.  An equivalent definition that flushes out the details of the foregoing is based on the vector ${\lambda}^{\downarrow}$ which is obtained from $\lambda$ by arranging the components of the latter in a nonincreasing order.  Then,  ${\lambda}^{1}\prec{\lambda}^{2}$ if ${\sum}_{i}^{j} {\lambda}^{1\downarrow}_{i} \leq {\sum}_{i}^{j} {\lambda}^{2\downarrow}_{i} $ for $j=1,2, \ldots d-1$, where $d$ is the larger of the two dimensions and trailing zeros are added where needed.  As stated earlier, the majorization relation is a \textit{partial} order, i.e., not every two vectors are comparable under majorization.  As suggested in \S I, two vectors are found to be incomparable when the difference in their degrees of uncertainty does not rise to the level required by majorization.  As evident from the second definition given above, majorization requires the satisfaction of $N-1$ inequalities if the number of non-zero components of the majorizing vector is $N$.  This accounts for the strength of the majorization relation as compared to scalar measures of disorder.  Indeed as alluded to in \S I, for any quasi-entropic function $F(\lambda)$, ${\lambda}^{1} \prec {\lambda}^{2}$ implies $F({\lambda}^{1}) \geq
F({\lambda}^{2})$, but not conversely.  On the other hand, if for \textit{every} quasi-entropic function $F(\lambda)$ we have
$F({\lambda}^{1}) \geq F({\lambda}^{2})$, then ${\lambda}^{1} \prec {\lambda}^{2}$  \cite{maj}.

To establish uncertainty bounds, we need to characterize the greatest lower bound, or \textit{infimum}, and the least upper bound, or \textit{supremum}, of a set of probability vectors \cite{HP3}.  The \textit{infimum} is defined as the vector that is majorized by every element of the set and in turn majorizes any vector with that property .   The \textit{supremum} is similarly defined.  We will briefly outline the construction of the \textit{infimum} here and refer the reader to Paper I for further details.

Given a set of probability vectors $\{ {\lambda}^{a} {\}}_{a=1}^{N}$, consider the vector ${\mu}^{inf}$ defined by
\begin{align}
{\mu}_{0}^{inf}=0&,\,\,\,{\mu}_{j}^{inf}=\min \big ( {\sum}_{i=1}^{j} {\lambda}^{1\downarrow}_{i},{\sum}_{i=1}^{j} {\lambda}^{2\downarrow}_{i},\ldots, \nonumber \\
&{\sum}_{i=1}^{j}{\lambda}^{N\downarrow}_{i} \big ), \,\,\, 1 \leq j \leq {d}_{max}, \label{1}
\end{align}
where ${d}_{max}$ is the largest dimension found in the set.  The infimum is then given by
\begin{equation}
{\lambda}^{inf}_{i}={[\inf({\lambda}^{1},{\lambda}^{2}, \ldots, {\lambda}^{N})]}_{i}={\mu}_{i}^{inf}-{\mu}_{i-1}^{inf}, \label{2}
\end{equation}
where $1 \leq i \leq {d}_{max}$.  The construction of the supremum starts with ${\mu}^{sup}$ in parallel with Eq.~(\ref{1}), but with ``max'' replacing ``min,'' and may require further steps detailed in Paper I.

It is worth noting here that the infimum (supremum) of a pair of probability vectors will in general be more (less) disordered than either.  Important special cases are (i) one of the two majorizes the other, in which case the latter is the infimum and the former the supremum of the two, and (ii) the two are equal, in which case either is both the infimum and the supremum.  Furthermore, a useful qualitative rule is that the more ``different'' are two probability vectors, the further will the infimum or supremum of the two be from at least one of them.  Finally, we note that while the infimum or supremum of a set of probability vectors always exists, it need not be a member of the set.
\subsection{Measurement and uncertainty}
A generalized measurement may be defined by a set of positive operators $\{ {\hat{\mathrm{E}}}_{\alpha} \}$ called measurement elements and subject to the completeness condition ${\sum}_{\alpha}{\hat{\mathrm{E}}}_{\alpha} = \hat{\mathbbm{1}}$.   The
probability that outcome $\alpha$ turns up in a measurement of the state $\hat{\rho}$ is given by the Born rule
$\mathscr{P}_{\alpha}(\hat{\rho})=\textrm{tr} [\hat{\mathrm{E}}_{\alpha} \hat{\rho} ]$.   A generalized measurement can always be considered to be the restriction of a more basic type, namely a \textit{projective} measurement, performed on an enlarged system to the system under generalized measurement \cite{NCH}.  A projective measurement is usually associated with an observable of the system represented by a self-adjoint operator $\hat{M}$, and entails a partitioning of the spectrum of $\hat{M}$ into a collection of subsets $\{ {b}_{\alpha}^{M} \}$ called measurement \textit{bins}.  We call a projective measurement \textit{maximal} if each bin consists of a single point of the spectrum of the measured observable.

Having assembled the necessary concepts, we can now characterize uncertainty by means of majorization relations in a natural manner.  To start, we define the probability vector $\mathscr{P}^{X}(\hat{\rho})$ resulting from a measurement $\mathrm{X}$ on a state $\hat{\rho}$ to be \textit{uncertain} if it is majorized by $\mathcal{I}=(1,0,\ldots,0)$ but not equal to it.  As such, $\mathscr{P}^{X}(\hat{\rho})$ is said to be strictly
majorized by $\mathcal{I}$ and written $\mathscr{P}^{X}(\hat{\rho})\prec \prec \mathcal{I}$. Similarly, given a pair of measurements $\mathrm{X}$ and $\mathrm{Y}$ on a state $\hat{\rho}$, we say $\mathscr{P}^{\mathrm{X}}(\hat{\rho})$ is more uncertain, equivalently more disordered, than $\mathscr{P}^{\mathrm{Y}}(\hat{\rho})$ if  $\mathscr{P}^{\mathrm{X}}(\hat{\rho}) \prec \mathscr{P}^{\mathrm{Y}}(\hat{\rho})$. Further, we define the joint uncertainty of a pair of measurements $\mathrm{X}$ and $\mathrm{Y}$ to be the outer product $\mathscr{P}^{\mathrm{X}}\otimes\mathscr{P}^{\mathrm{Y}}$, i.e., $\mathscr{P}_{\alpha
\beta}^{\mathrm{X} \oplus \mathrm{Y}}=\mathscr{P}^{\mathrm{X}}_{\alpha} \mathscr{P}^{\mathrm{Y}}_{\beta}$.  Since
$H(\mathscr{P}^{\mathrm{X}}\otimes \mathscr{P}^{\mathrm{Y}})=H(\mathscr{P}^{\mathrm{X}})+H(\mathscr{P}^{\mathrm{Y}})$, where $H(\cdot)$ is the Shannon entropy function, this definition is seen to be consistent with its entropic counterpart.   As stated earlier, $\mathscr{P}^{\mathrm{X}} \prec \mathscr{P}^{\mathrm{Y}}$ implies $H(\mathscr{P}^{\mathrm{X}}) \geq H(\mathscr{P}^{\mathrm{Y}})$ but not conversely.  These definitions naturally extend to an arbitrary number of states and measurements.

We are now in a position to state the majorization statement of the uncertainty principle established in Paper I:

``The joint results of a set of generalized measurements of a given state are no less uncertain than a probability vector that depends on the measurement set but not the state, and is itself uncertain unless the measurement elements have a common eigenstate."

In symbols,
\begin{equation}
\mathscr{P}^{\mathrm{X}}(\hat{\rho})\otimes \mathscr{P}^{\mathrm{Y}}(\hat{\rho}) \otimes \ldots \otimes \mathscr{P}^{\mathrm{Z}}(\hat{\rho}) \prec
{\mathscr{P}}_{sup}^{\mathrm{X} \oplus \mathrm{Y} \oplus \ldots \oplus \mathrm{Z}} \prec \prec \mathcal{I}, \label{3}
\end{equation}
where
\begin{equation}
{\mathscr{P}}_{sup}^{\mathrm{X} \oplus \mathrm{Y} \oplus \ldots \oplus \mathrm{Z}}=
{\sup}_{\hat{\rho}} [\mathscr{P}^{\mathrm{X}}(\hat{\rho})\otimes \mathscr{P}^{\mathrm{Y}}(\hat{\rho}) \otimes \ldots \otimes
\mathscr{P}^{\mathrm{Z}}(\hat{\rho})] , \label{4}
\end{equation}
unless the measurement elements $\{ \hat{\mathrm{E}}^{\mathrm{X}}, \hat{\mathrm{E}}^{\mathrm{Y}}, \ldots, \hat{\mathrm{E}}^{\mathrm{Z}} \}$ have a common eigenstate in which case ${\mathscr{P}}_{sup}^{\mathrm{X} \oplus \mathrm{Y} \oplus \ldots \oplus \mathrm{Z}} = \mathcal{I}$.

Note that ${\mathscr{P}}_{sup}^{\mathrm{X} \oplus \mathrm{Y} \oplus \ldots \oplus \mathrm{Z}}$ is the majorization uncertainty bound for the measurement set considered.

\subsection{majorization uncertainty bounds}

As stated above, the uncertainty bound ${\mathscr{P}}_{sup}^{\mathrm{X} \oplus \mathrm{Y} \oplus \ldots \oplus \mathrm{Z}}$ given by Eq.~(\ref{4}) depends on the measurement set but not the state of the system.  As the supremum of all possible measurement outcomes, it is the probability vector that sets the irreducible lower bound to uncertainty for the set.  As such, it is the counterpart of the variance product or entropic lower bound in the traditional formulations of the uncertainty principle.  Unlike the latter, however, ${\mathscr{P}}_{sup}^{\mathrm{X} \oplus \mathrm{Y} \oplus \ldots \oplus \mathrm{Z}}$ is in general not realizable or even approachable by any state of the system.  Therefore, there is in general no such thing as a ``minimum uncertainty state'' within the majorization framework.  This is a consequence of the fact, mentioned earlier, that the infimum or supremum of a set of vectors need not be a member of the set.  An obvious special case is the trivial example of zero uncertainty for which ${\mathscr{P}}_{sup}^{\mathrm{X} \oplus \mathrm{Y} \oplus \ldots \oplus \mathrm{Z}}= \mathcal{I}$, signaling the existence of a common eigenstate for the measurement elements.

Intuitively, we expect that mixing states can only increase their uncertainty, as is known to be the case for scalar measures of uncertainty.  In case of majorization, this expectation is manifested in the property that the uncertainty bound ${\mathscr{P}}_{sup}^{\mathrm{X} \oplus \mathrm{Y} \oplus \ldots \oplus \mathrm{Z}}$ is in general realized on the class of pure states.  Indeed for two and three mutually unbiased observables on a two-dimensional Hilbert space considered, we found in Paper I that the required maxima for the components of ${\mu}^{sup}$ that serve to define the uncertainty bound are reached on pure states.  More specifically, as outlined in \S IIA above and detailed in \S IIB, IVA and IVB of Paper I, the calculation of ${\mathscr{P}}_{sup}^{\mathrm{X} \oplus \mathrm{Y} \oplus \ldots \oplus \mathrm{Z}}$ involves a component-wise determination of ${\mu}^{sup}$ by a series of maximizations over all possible density matrices $\hat{\rho}$.  The property in question then guarantees that the maximization process can be limited to density matrices representing pure states only.  We will prove this assertion in general, as well as for the class of separable states, in the Appendix.

\section{Entanglement Detection}
As discussed earlier, an entanglement detector can be effective in deciding whether a given density matrix is separable by providing a condition that is satisfied by all separable states and if violated signals entanglement.   A large body of entanglement detection strategies have been developed in recent years which are primarily based on scalar conditions, including those based on variance and entropic type uncertainty relations, which can be found in Refs. \cite{sep,HOR,GIO,GAM}.  Here we shall develop majorization conditions for entanglement detection based on the formulation of measurement uncertainty given in Paper I and outlined in \S II above.  In particular, the entanglement condition given in Theorem 1 below is linear and susceptible to experimental implementation, so it can be formulated as an entanglement witness.   Nonlinear detectors developed below, on the other hand, rely on majorization-based uncertainty relations.  We will also introduce the notion of subsystem disorder and a sharpened version of the Nielsen-Kempe \cite{nik} separability condition as a nonlinear entanglement detector.

As majorization relations, our results in general entail multiple inequalities whose number will grow with the uncertainty levels involved.  As discussed in \S IIA, this is an important feature of the majorization relation as comparator of disorder, one that sets it apart from scalar conditions and provides for the refinement needed in comparing highly disordered vectors.  As will be seen in \S IV, this property has the consequence that each majorization condition yields a scalar condition for the entire quasi-entropic class of uncertainty measures.

\subsection{Linear detectors}
Our detection strategy is based on the intuitive expectation that the measurement uncertainty bound for the class of separable states of a multipartite system must be majorized by the corresponding bound for all states, and that this hierarchy can be exploited for entanglement detection.  This is the majorization rendition of the strategy often used to derive separability conditions \cite{GIO,GAM}.  We will first consider the case of one generalized measurement resulting in a linear detection condition.

\textbf{Theorem 1.} Let the results of the generalized measurement ($\mathrm{X}, \{ \hat{\mathrm{E}}_{\alpha}^{X} \}$) on a multipartite state ${\hat{\rho}}^{ABC \dots F}$ of parties $(A, B, C, \dots, F)$ defined on a finite-dimensional Hilbert space be bounded by
\begin{equation}
{\mathscr{P}}_{sup}^{\mathrm{X}}={\sup}_{{\hat{\rho}}^{ABC \dots F}}{\mathscr{P}}^{\mathrm{X}}({\hat{\rho}}^{ABC \dots F}), \label{5}
\end{equation}
and in the case of a separable state ${\hat{\rho}}^{ABC \dots F}_{sep}$ by
\begin{equation}
{\mathscr{P}}_{sep;sup}^{\mathrm{X}}={\sup}_{{\hat{\rho}}^{ABC \dots F}_{sep}}{\mathscr{P}}^{\mathrm{X}}({\hat{\rho}}^{ABC \dots F}_{sep}). \label{6}
\end{equation}
Then, (i) the suprema in the foregoing pair of equations may be taken over the class of pure and pure, product states, respectively, and (ii) given an arbitrary state ${\hat{\sigma}}^{ABC \dots F}$, the condition ${\mathscr{P}}^{\mathrm{X}}({\hat{\sigma}}^{ABC \dots F}) \nprec {\mathscr{P}}_{sep;sup}^{\mathrm{X}}$ implies that ${\hat{\sigma}}^{ABC \dots F}$ is entangled \cite{noteent}.

Part (i) of Theorem 1 is the majorization version of the familiar result that uncertainty bounds are realized on pure states and is proved in the Appendix \cite{noteext}.   Part (ii) is the statement that the bound in Eq.~(\ref{6}) is  a necessary condition for separability and its violation signals the existence of entanglement.  As noted above, Theorem 1 provides an operational method of entanglement detection and can be reformulated as a set of  entanglement witnesses.

Clearly, the efficacy of Theorem 1 in detecting entanglement depends on the choice of the measurement ${\mathrm{X}}$ for a given quantum state.  This suggests looking for the optimum measurement, in parallel with the optimization problem for entanglement witnesses \cite{LKCH}.  Note that here the choice of the generalized measurement ${\mathrm{X}}$ affects both ${\mathscr{P}}^{\mathrm{X}}({\hat{\sigma}}^{ABC \dots F})$ and ${\mathscr{P}}_{sep;sup}^{\mathrm{X}}$, thus making the task of finding a fully optimized solution rather difficult.   It is therefore fortunate that we can achieve a partial optimization on the basis of Theorem 3 of Paper I, the relevant content of which we can state as follows:

\textbf{Lemma} Under the conditions of Theorem 1 and with ${\mathrm{X}}$ restricted to rank 1 measurements, we have
\begin{equation}
{\mathscr{P}}^{\mathrm{X}}({\hat{\sigma}}^{ABC \dots F}) \prec {\mathscr{P}}^{{\mathrm{X}}^{\star}}({\hat{\sigma}}^{ABC \dots F})={\lambda}({\hat{\sigma}}^{ABC \dots F}), \label{7}
\end{equation}
where ${\mathrm{X}}^{\star}$ is a maximal projective measurement whose elements are the set of rank 1 orthogonal projection operators onto the eigenvectors of ${\hat{\sigma}}^{ABC \dots F}$ and ${\lambda}({\hat{\sigma}}^{ABC \dots F})$ is the corresponding spectrum.

It should be noted here that while the choice of ${{\mathrm{X}}^{\star}}$ succeeds in optimizing ${\mathscr{P}}^{\mathrm{X}}({\hat{\sigma}}^{ABC \dots F})$ which appears in part (ii) of Theorem 1, its effect on ${\mathscr{P}}_{sep;sup}^{\mathrm{X}}$, the other object appearing therein, is unknown and may result in a poor overall choice.  Moreover, any degeneracy in the spectrum of ${\hat{\sigma}}^{ABC \dots F}$ renders ${{\mathrm{X}}^{\star}}$ nonunique and subject to further optimization.   These features will be at play in the example considered below where, the foregoing caveats notwithstanding, the resulting detector turns out to be fully optimal.

The example in question is the Werner state ${\hat{\rho}}^{wer}_{d}(q)$ for a bipartite system of two $d$-level parties $A$ and $B$ defined on a $d^2$-dimensional Hilbert space \cite{wer,pitr}
\begin{equation}
{\hat{\rho}}^{wer}_{d}(q)=\frac{1}{d^2}({1-q})\hat{{\mathbbm{1}}} + q \mid{\mathfrak{B}}_{1} \rangle \langle {\mathfrak{B}}_{1}\mid,  \label{8}
\end{equation}
where
\begin{equation}
 \mid{\mathfrak{B}}_{1} \rangle \ =\frac{1}{\sqrt{d}} {\sum}_{j=0}^{d-1} \mid A,j \rangle \otimes \mid B,j \rangle. \label{9}
\end{equation}
Here $\{ \mid A,j \rangle \}$  and $\{ \mid B,j \rangle \}$) are orthonormal bases for subsystem $A$ and $B$ respectively.  Also, here and elsewhere we use the symbol $\hat{{\mathbbm{1}}}$ to denote the identity operator of dimension appropriate to the context.  Note that $\mid{\mathfrak{B}}_{1} \rangle$ is the generalization of the two-qubit Bell state $(\mid 00 \rangle + \mid 11 \rangle )/\sqrt{2}$.  As such, it is totally symmetric, as well as maximally entangled in the sense that its marginal states are maximally disordered and equal to $\hat{{\mathbbm{1}}}/d$. An important fact to be used in the following is that ${\hat{\rho}}^{wer}_{d}(q)$ is separable if and only if $q \leq (1+d)^{-1}$ \cite{pitr}.

In order to probe the separability of ${\hat{\rho}}^{wer}_{d}(q)$ by means of Theorem 1, we must first choose a measurement.  We will use the above Lemma to guide our choice of the putative optimum measurement ${\mathrm{X}}^{\star}$.  The Lemma requires that the measurement elements be equal to the orthogonal projections onto the eigenvectors of ${\hat{\rho}}^{wer}_{d}(q)$.   However, a simple calculation shows that the spectrum of ${\hat{\rho}}^{wer}_{d}(q)$ is degenerate, consisting of a simple eigenvalue equal to $q+{d}^{-2}(1-q)$ and a $({d}^{2}-1)$-fold degenerate eigenvalue equal to ${d}^{-2}(1-q)$.  A natural choice of basis for the degenerate subspace is a set of $d^2-1$ generalized Bell states which, together with $\mid{\mathfrak{B}}_{1} \rangle $ of Eq.~(\ref{9}), constitute the orthonormal eigenbasis ${\{\mid{\mathfrak{B}}_{\alpha}  \rangle \}}_{\alpha=1}^{d^2}$ for ${\hat{\rho}}^{wer}_{d}(q)$.  This generalized Bell basis is thus characterized by the fact that each $\mid{\mathfrak{B}}_{\alpha}  \rangle$ is maximally entangled and possesses equal Schmidt coefficients.

The measurement elements of the optimal measurement ${\mathrm{X}}^{\star}$ are thus given by ${\hat{\mathrm{E}}}_{\alpha}^{{X}^{\star}}= \mid {\mathfrak{B}}_{\alpha}  \rangle  \langle {\mathfrak{B}}_{\alpha} \mid $, $\alpha=1, 2, \dots, d^2$.  With ${\mathrm{X}}^{\star}$ so defined, we readily find
\begin{align}
{\mathscr{P}}^{{\mathrm{X}}^{\star}}[{\hat{\rho}}^{wer}_{d}(q)]=&[q+{d}^{-2}(1-q), {d}^{-2}(1-q), \nonumber \\
&{d}^{-2}(1-q), \ldots, {d}^{-2}(1-q)], \label{10}
\end{align}
which corresponds to the spectrum of ${\hat{\rho}}^{wer}_{d}(q)$ as expected.  Note that the Lemma guarantees that any other rank 1 measurement in place of ${\mathrm{X}}^{\star}$ would produce a more disordered probability vector on the right-hand side of Eq.~(\ref{10}).

The next step is the calculation of ${\mathscr{P}}_{sep;sup}^{{\mathrm{X}}^{\star}}$ of Theorem 1, which is the least disordered probability vector that can result from the measurement of ${\mathrm{X}}^{\star}$ on a pure, product state of two $d$-level subsystems.  As can be seen in Eq.~(\ref{A7}) of the Appendix, this calculation requires finding the maximum overlap of every measurement element $\mid {\mathfrak{B}}_{\alpha}  \rangle  \langle {\mathfrak{B}}_{\alpha} \mid $ with the class of pure, product states.   This overlap is given by the square of the largest Schmidt coefficient of the respective Bell state $\mid {\mathfrak{B}}_{\alpha}\rangle$ \cite{MBetal}.  Since all Schmidt coefficients are equal to $1/\sqrt{d}$ for every Bell state $\mid {\mathfrak{B}}_{\alpha}\rangle$, we can conclude that
\begin{equation}
{\mathscr{P}}_{sep;sup}^{{\mathrm{X}}^{\star}}=(1/d, 1/d, \ldots, 1/d,0,0, \ldots,0).  \label{11}
\end{equation}

We are now in a position to apply the entanglement criterion of Theorem 1, which reads ${\mathscr{P}}^{\mathrm{{X}^{\star}}}[{\hat{\rho}}^{wer}_{d}(q)] \nprec {\mathscr{P}}_{sep;sup}^{\mathrm{{X}^{\star}}}$ in this instance \cite{rem}.  Using the information in Eqs.~(\ref{10}) and (\ref{11}), we readily see that this criterion translates to the single inequality $q+{d}^{-2}(1-q) > 1/d$, or $q > 1/(1+d)$, which is the necessary and sufficient condition for the inseparability of the Werner state ${\hat{\rho}}^{wer}_{d}(q)$.  Thus every entangled Werner state of two $d$-level systems is detected by the measurement ${\mathrm{X}}^{\star}$ defined above.

While the efficacy of the generalized Bell states for detecting entanglement in Werner states is well established \cite{sep,HOR,GIO,GAM}, we note its natural emergence from the general majorization results of this subsection.   It should also be noted that Theorem 1 and the foregoing analysis can be extended to multipartite states.

\subsection{Nonlinear detectors}
As an example of nonlinear entanglement detectors, we will derive majorization conditions for separability based on uncertainty relations.  The method we will follow relies on the degeneracy properties of projective measurements on a single system versus the products of such measurements on a multipartite system \cite{GIO, GAM}.  For example, with measurement ${\mathrm{X}}^{A}$ on ${\hat{\rho}}^{A}$ and ${\mathrm{X}}^{B}$ on ${\hat{\rho}}^{B}$ and each having a simple spectrum, the product projective measurement ${\mathrm{X}}^{A} \otimes {\mathrm{X}}^{B}$ on the bipartite system ${\hat{\rho}}^{AB}$ may be degenerate.  A specific case is ${{\hat{\sigma}}_{x}}^{A}\otimes {{\hat{\sigma}}_{x}}^{B}$, whose spectrum $(+1/4,+1/4,-1/4,-1/4)$ is doubly degenerate, versus ${{\hat{\sigma}}_{x}}^{A}$ or ${{\hat{\sigma}}_{x}}^{B}$ each of which has the simple spectrum $(+1/2,-1/2)$.  Note the fact that here we are following common practice by defining ${{\hat{\sigma}}_{x}}^{A}\otimes {{\hat{\sigma}}_{x}}^{B}$ to be the projective measurement whose two elements project into the subspaces corresponding to the eigenvalues $+1/4$ and $-1/4$ (and not the rank 1 projective measurement that resolves the degeneracies by virtue of using all four product elements).  Consequently, it may happen that ${\mathrm{X}}^{A}\otimes {\mathrm{X}}^{B}$ and ${\mathrm{Y}}^{A}\otimes {\mathrm{Y}}^{B}$ have a common eigenstate while ${\mathrm{X}}^{A}$ and ${\mathrm{Y}}^{A}$ do not, and that the said common eigenstate is an entangled pure state.  In such cases, the corresponding probability vectors will reflect the stated differences and may be capable of detecting entanglement as in the case of linear detectors.

Consider two product projective measurements ${\mathrm{X}}^{A} \otimes {\mathrm{X}}^{B}$ and ${\mathrm{Y}}^{A}\otimes {\mathrm{Y}}^{B}$ performed on the bipartite state ${\hat{\rho}}^{AB}$.  The measurement results are then given by
\begin{align}
{\mathscr{P}}^{{\mathrm{X}}^{A}\otimes {\mathrm{X}}^{B}}({\hat{\rho}}^{AB})= \textrm{tr}&[{\hat{\Pi}}^{{\mathrm{X}}^{A}\otimes {\mathrm{X}}^{B}} {\hat{\rho}}^{AB} ], \nonumber \\ {\mathscr{P}}^{{\mathrm{Y}}^{A}\otimes {\mathrm{Y}}^{B}}({\hat{\rho}}^{AB})= \textrm{tr}&[{\hat{\Pi}}^{{\mathrm{Y}}^{A}\otimes {\mathrm{Y}}^{B}} {\hat{\rho}}^{AB} ], \nonumber \\
\mathscr{P}^{ ({\mathrm{X}}^{A}\otimes {\mathrm{X}}^{B}) \oplus ({\mathrm{Y}}^{A}\otimes {\mathrm{Y}}^{B})}({\hat{\rho}}^{AB}) &=\mathscr{P}^{ {\mathrm{X}}^{A} \otimes {\mathrm{X}}^{B} }({\hat{\rho}}^{AB})
\nonumber \\
\otimes \mathscr{P}^{{\mathrm{Y}}^{A} \otimes {\mathrm{Y}}^{B} }({\hat{\rho}}^{AB}),  \label{12}
\end{align}
where ${\hat{\Pi}}^{{\mathrm{X}}^{A}\otimes {\mathrm{X}}^{B}}$ and ${\hat{\Pi}}^{{\mathrm{Y}}^{A}\otimes {\mathrm{Y}}^{B}}$, both projection operators but not necessarily rank 1, represent the measurement elements of the two product measurements.  Note the majorization definition of joint uncertainty, given earlier, at work on the last line of Eq.~(\ref{12}).
Under these conditions, we have the following general result.

\textbf{Theorem 2}.  The results of measurements ${\mathrm{X}}^{A}\otimes {\mathrm{X}}^{B}$ and ${\mathrm{Y}}^{A}\otimes
{\mathrm{Y}}^{B}$ on a separable state ${\hat{\rho}}^{AB}_{sep}$ satisfy
\begin{equation}
\mathscr{P}^{ ({\mathrm{X}}^{A}\otimes {\mathrm{X}}^{B}) \oplus ({\mathrm{Y}}^{A}\otimes {\mathrm{Y}}^{B})}({\hat{\rho}}^{AB}_{sep}) \prec
\mathscr{P}_{sup}^{ {\mathrm{X}}\oplus {\mathrm{Y}}},  \label{13}
\end{equation}
where $\mathscr{P}_{sup}^{ {\mathrm{X}}\oplus {\mathrm{Y}}}$ is defined in Eq.~(\ref{4}), hence the condition $\mathscr{P}^{ ({\mathrm{X}}^{A}\otimes {\mathrm{X}}^{B}) \oplus ({\mathrm{Y}}^{A}\otimes {\mathrm{Y}}^{B})}({\hat{\sigma}}^{AB}) \nprec \mathscr{P}_{sup}^{ {\mathrm{X}}\oplus {\mathrm{Y}}}$ implies that the state ${\hat{\sigma}}^{AB}$ is entangled.

To establish Theorem 2, consider product states of the form ${\hat{\rho}}^{A} \otimes {\hat{\rho}}^{B}$.  Then Lemma 1 of Ref.~\cite{GAM} asserts that
\begin{align}
\mathscr{P}^{ {\mathrm{X}}^{A}\otimes {\mathrm{X}}^{B} }({\hat{\rho}}^{A} \otimes {\hat{\rho}}^{B}) &\prec \mathscr{P}^{ {\mathrm{X}}^{A}}({\hat{\rho}}^{A}), \nonumber \\
\mathscr{P}^{ {\mathrm{Y}}^{A}\otimes {\mathrm{Y}}^{B} }({\hat{\rho}}^{A} \otimes {\hat{\rho}}^{B}) &\prec \mathscr{P}^{ {\mathrm{Y}}^{A}}({\hat{\rho}}^{A}). \label{14}
\end{align}
Using these relations and Eq.~(\ref{12}), we find \cite{crs}
\begin{equation}
\mathscr{P}^{ ({\mathrm{X}}^{A}\otimes {\mathrm{X}}^{B}) \oplus ({\mathrm{Y}}^{A}\otimes {\mathrm{Y}}^{B})}({\hat{\rho}}^{A} \otimes {\hat{\rho}}^{B}) \prec
\mathscr{P}^{ {\mathrm{X}}^{A}}({\hat{\rho}}^{A})\otimes \mathscr{P}^{ {\mathrm{Y}}^{A}}({\hat{\rho}}^{A}).  \label{15}
\end{equation}
Since the right-hand side of Eq.~(\ref{15}) is by definition majorized by $\mathscr{P}_{sup}^{ {\mathrm{X}}\oplus
{\mathrm{Y}}}$, we conclude that
\begin{equation}
\mathscr{P}^{ ({\mathrm{X}}^{A}\otimes {\mathrm{X}}^{B}) \oplus ({\mathrm{Y}}^{A}\otimes {\mathrm{Y}}^{B})}({\hat{\rho}}^{A} \otimes {\hat{\rho}}^{B}) \prec \mathscr{P}_{sup}^{ {\mathrm{X}}\oplus {\mathrm{Y}}}.  \label{16}
\end{equation}

At this point we note that $\mathscr{P}^{ ({\mathrm{X}}^{A}\otimes {\mathrm{X}}^{B}) \oplus ({\mathrm{Y}}^{A}\otimes {\mathrm{Y}}^{B})}_{sep;sup}$, which by definition majorizes all probability vectors that can appear on the left-hand side of Eq.~(\ref{13}), can be found among pure, product states \cite{noteext}.  This assertion is established in Eq.~(\ref{A8}) et seq. of the Appendix.  Consequently, the product state ${\hat{\rho}}^{A} \otimes {\hat{\rho}}^{B}$ in Eq.~(\ref{16}) may be replaced by any separable state ${\hat{\rho}}^{AB}_{sep}$, thereby establishing Eq.~(\ref{13}) and Theorem 2.

We note in passing that, because the measurements considered in Theorem 2 are in general not rank 1, the Lemma which we used earlier for detector optimization cannot be applied here.

To illustrate Theorem 2, we will consider the case of three mutually unbiased observables measured on bipartite states of two-level systems, as in \S IVB of Paper I.  In effect, this amounts to measuring products of the three spin components of a pair of spin-1/2 systems (or qubits).  For this case, Eq.~(\ref{16}) reads
\begin{equation}
\mathscr{P}^{ ({{\sigma}_{x}}^{A}\otimes {{\sigma}_{x}}^{B}) \oplus
({{\sigma}_{y}}^{A}\otimes {{\sigma}_{y}}^{B})\oplus ({{\sigma}_{z}}^{A}\otimes {{\sigma}_{z}}^{B})}({\hat{\rho}}^{AB}_{sep})\prec \mathscr{P}_{sup}^{
{\sigma}_{x}\oplus {\sigma}_{y}\oplus {\sigma}_{z}},  \label{17}
\end{equation}
where ${\hat{\rho}}^{AB}_{sep}$ is any separable state of two qubits.  Thus a violation of this relation by a two-qubit state implies that it is entangled.

For the set of states ${\hat{\rho}}^{AB}$ to be probed by Theorem 2, we will consider the Werner family of two-qubit states
\begin{equation}
{\hat{\rho}}^{wer}(q)=\frac{1}{4}({1-q}){\mathbbm{1}}+ q \mid{\mathfrak{B}}_{1} \rangle \langle {\mathfrak{B}}_{1}  \mid,  \label{18}
\end{equation}
which is known to be entangled if and only if $q >1/3$.  Here $0 \leq q \leq 1$ and $\mid{\mathfrak{B}}_{1} \rangle =(\mid 00 \rangle + \mid 11 \rangle )/\sqrt{2}$ is the totally symmetric Bell state of Eq.~(\ref{9}) for the present case.

A calculation of
the probability vector for this measurement gives $(1 \pm q)(1 \pm q)(1 \pm q)/8$ for the $8$ components of $\mathscr{P}^{
({{\sigma}_{x}}^{A}\otimes {{\sigma}_{x}}^{B}) \oplus ({{\sigma}_{y}}^{A}\otimes {{\sigma}_{y}}^{B})\oplus
({{\sigma}_{z}}^{A}\otimes {{\sigma}_{z}}^{B})}[{\hat{\rho}}^{wer}(q)]$.  Arranged in a non-ascending order, these are ${(1+q)}^{3}/8$, ${(1+q)}^{2}{(1-q)}/8$ and ${(1+q)}{(1-q)}^{2}/8$ both three-fold degenerate, and ${(1-q)}^{3}/8$.   The supremum $\mathscr{P}_{sup}^{
{\sigma}_{x}\oplus {\sigma}_{y}\oplus {\sigma}_{z}}$, on the other hand, was calculated in \S IVB of Paper I, where it was found as
\begin{align}
\mathscr{P}_{sup}^{
{\sigma}_{x}\oplus {\sigma}_{y}\oplus {\sigma}_{z}}=\frac{1}{8}\big[{(1+1/\sqrt{3})}^{3},2{(1+1/\sqrt{2})}^{2} \nonumber \\
-{(1+1/\sqrt{3})}^{3}, 4-{(1+1/\sqrt{2})}^{2}, 4-{(1+1/\sqrt{2})}^{2}\nonumber \\
 ,0,0,0,0 \big].  \label{19}
\end{align}

According to Theorem 2, ${\hat{\rho}}^{wer}(q)$ violates the separability condition of Eq.~(\ref{13}) if $\mathscr{P}_{sup}^{
{\sigma}_{x}\oplus {\sigma}_{y}\oplus {\sigma}_{z}}$ fails to majorize $\mathscr{P}^{
({{\sigma}_{x}}^{A}\otimes {{\sigma}_{x}}^{B}) \oplus ({{\sigma}_{y}}^{A}\otimes {{\sigma}_{y}}^{B})\oplus
({{\sigma}_{z}}^{A}\otimes {{\sigma}_{z}}^{B})}[{\hat{\rho}}^{wer}(q)]$, which occurs for $q > 1/\sqrt{3}=0.577$ in this example.   The entanglement in ${\hat{\rho}}^{wer}(q)$ is thus detected by Theorem 2 for $q > 0.577$, and missed for the range $0.333 < q \leq 0.577$.  By comparison, a similar method based on the Shannon entropy in Ref. \cite{GIO} detects entanglement in ${\hat{\rho}}^{wer}(q)$ for $q >0.65$.   Reference \cite{GAM}, using the same method but relying on the family of Tsallis entropies, matches our condition $q > 0.577$ by means of a numerical calculation that searches over the Tsallis family for optimum performance.

We end this subsection by a brief description of a sharpened version of the Nielsen-Kempe theorem as a nonlinear entanglement detector \cite{HP3}.  This celebrated theorem is an elegant separability condition based on majorization relations.  It asserts that the spectrum of a separable bipartite state is majorized by each of its marginal spectra.  By extension, this theorem guarantees that the spectra of all possible subsystems of a separable multipartite system must majorize its global spectrum \cite{nik}.   Our version of this theorem employs the notion of the infimum of a set of probability vectors discussed in \S IIA, and is based on the observation that if a probability vector is majorized by each member of a set of probability vectors, then it must also be majorized by the infimum of that set.

Consider the multipartite state ${\hat{\rho}}^{ABC \dots F}$ together with all its subsystem states ${\{ {\hat{\rho}}^{{X}_{a}} \}}_{a=1}^{f}$ and subsystem spectra ${\{ {\lambda}^{{X}_{a}} \}}_{a=1}^{f}$, where ${\{ {{X}_{a}} \}}_{a=1}^{f}$ represent all proper subsets of the set of parties $(A, B, C,  \dots, F)$.  Then the infimum of the subsystem spectra, ${\Lambda}^{ABC \dots F}={\inf}[{\lambda}^{{X}_{1}},{\lambda}^{{X}_{2}}, \ldots, {\lambda}^{{X}_{f}}]$, embodies the \textit{subsystem disorder} of the state ${\hat{\rho}}^{ABC \dots F}$, as expressed in the following theorem.

\textbf{Theorem 3} (Nielsen-Kempe).  A multipartite state is entangled if its system disorder fails to exceed its subsystem disorder, i.e., ${\lambda}^{ABC \dots F}$ is entangled if ${\lambda}^{ABC \dots F}\nprec {\Lambda}^{ABC \dots F}$.

\section{Quasi-Entropic Detectors}
The majorization based theorems of the previous section have direct corollaries that yield scalar entanglement detectors for the entire class of \textit{Schur-concave} measures.   A Schur-concave function $G$ is defined by the property that ${\lambda}^{1} \prec {\lambda}^{2}$ implies $G({\lambda}^{1}) \geq G({\lambda}^{2})$ \cite{ftn1}.   Equivalently, Schur-concave functions are characterized by being monotonic with respect to the majorization relation.  They include all functions $G(\cdot)$ that are concave and symmetric with respect to the arguments, a subset which we have defined as quasi-entropic.  An important subclass of quasi-entropic functions is obtained if we restrict $G$ to have the following trace structure:
\begin{equation}
G(\lambda)=\mathrm{tr}[g(\lambda)]= {\sum}_{\alpha} g({\lambda}_{\alpha}), \label{20}
\end{equation}
where $g(\cdot)$ is a concave function of a single variable and $\lambda$ is treated as a diagonal matrix for the purpose of calculating the trace.  Note that $G(\cdot)$ as constructed in Eq.~(\ref{20}) is manifestly symmetric and, as a sum of concave functions, it is also concave.   We shall refer to this class of functions, which include the Shannon and Tsallis entropies, as \textit{trace type} quasi-entropic.  An important fact regarding this class of measures is that if, for a given pair of probability vectors $\lambda$ and $\mu$, we have $G({\lambda})>G({\mu})$ for every trace type quasi-entropic measure $G$, then $\lambda \prec \mu$.

The standard entropic measure of uncertainty \cite{DEU}, which is based on the Shannon entropy function, is trace type quasi-entropic and corresponds to the choice $g(x)=H(x)=-x \ln(x)$ in Eq.~(\ref{20}).  An example of an information theoretically relevant measure that is quasi-entropic but not trace type is the R\'{e}nyi subfamily of entropies of order less than one \cite{ftn2}.   Accordingly, although the scalar entanglement detectors that follow from the theorems of \S III actually hold for the entire class of Schur-concave measures, we will continue to focus on the quasi-entropic class as the most suitable for information theoretical applications.

Scalar entanglement detectors based on Shannon, Tsallis, and R\'{e}nyi entropies are already well known, albeit with detection bounds that may differ from those reported here \cite{GIO,GAM,sep}.   Our primary purpose here is to emphasize the natural and categorical manner in which majorization based entanglement detectors yield entire classes of scalar detectors.

With $G(\cdot)$ a quasi-entropic function, the aforementioned monotonicity property implies the following corollaries to Theorems 1-3.\\

\noindent
\textbf{Corollary 1}. Under the conditions of Theorem 1, the multipartite state ${\hat{\sigma}}^{ABC \dots F}$ is entangled if
\begin{equation}
G[{\mathscr{P}}^{\mathrm{X}}({\hat{\sigma}}^{ABC \dots F})] <  G[{\mathscr{P}}_{sep;sup}^{\mathrm{X}}].  \label{21}
\end{equation}
\textbf{Corollary 2}. Under the conditions of Theorem 2, the bipartite state ${\hat{\sigma}}^{AB}$ is entangled if
\begin{equation}
G[\mathscr{P}^{ ({\mathrm{X}}^{A}\otimes {\mathrm{X}}^{B})}({\hat{\sigma}}^{AB})]+ G[\mathscr{P}^{ ({\mathrm{Y}}^{A}\otimes {\mathrm{Y}}^{B})} ({\hat{\sigma}}^{AB})] < G[\mathscr{P}_{sup}^{ {\mathrm{X}}\oplus {\mathrm{Y}}}]. \label{22}
\end{equation}
\textbf{Corollary 3}. Under the conditions of Theorem 3, the multipartite state ${\hat{\sigma}}^{ABC \dots F}$ is entangled if
\begin{equation}
G[{\lambda}^{ABC \dots F}] < G[{\Lambda}^{ABC \dots F}]. \label{23}
\end{equation}

As an application of the above corollaries, we will consider the entanglement detection threshold for the bipartite Werner state considered in Eq.~(\ref{8}) et seq.   Using Corollary 1 together with the Tsallis entropy function, we find that ${\hat{\rho}}^{wer}_{d}(q)$ is entangled if
\begin{equation}
{S}^{tsa}_{r}\big[ {\mathscr{P}}^{{\mathrm{X}}^{\star}}[{\hat{\rho}}^{wer}_{d}(q)] \big ] < {S}^{tsa}_{r}\big[ {\mathscr{P}}_{sep;sup}^{{\mathrm{X}}^{\star}} \big ], \label{24}
\end{equation}
where ${S}^{tsa}_{r}(\cdot)$ is the Tsallis entropy of order $r$, and the two probability vectors appearing in Eq.~(\ref{24}) are those in Eqs.~(\ref{10}) and (\ref{11}).   The Tsallis entropy function ${S}^{tsa}_{r}(\cdot)$ corresponds to the choice $g(\lambda)= (\lambda-{\lambda}^{r})/(r-1)$, $1<r<\infty$, in Eq.~(\ref{20}).   The Tsallis entropy of order 1 is defined by continuity and equals the Shannon entropy \cite{tsa}.

\begin{figure}
\includegraphics[]{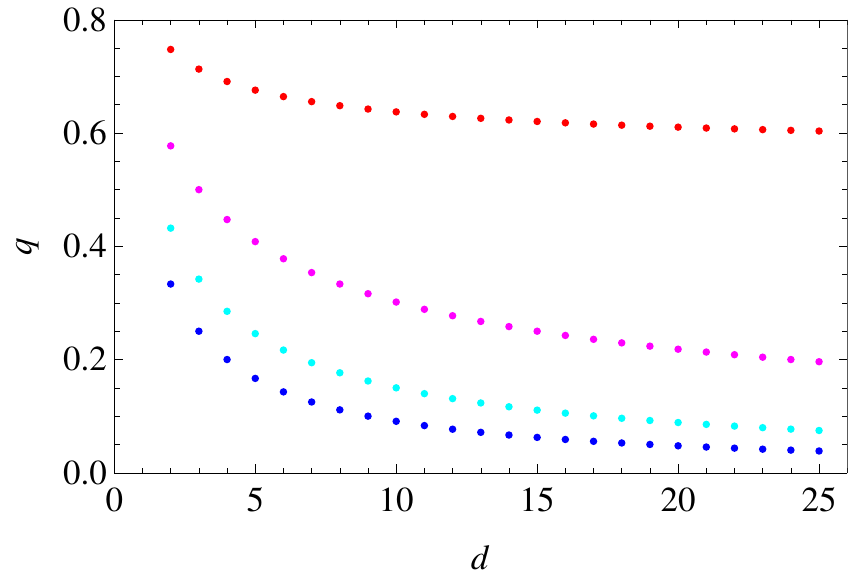}
\caption{Entanglement detection threshold $q$ for a bipartite Werner state versus dimension $d$ of each party, using Corollary 1 and the Tsallis entropy.  The order of the Tsallis entropy equals, from top to bottom respectively, 1(red), 2 (magenta), 5 (turquoise), and $\infty$ (blue).  The top graph corresponds to the Shannon entropy and the bottom one (d) reproduces the known entanglement threshold for the state.}
\label{fig1}
\end{figure}

A straightforward calculation turns Eq.~(\ref{24}) into
\begin{align}
&\frac{1-{[q+(1-q)/{d}^{2})]}^{r}-({d}^{2}-1){[(1-q)/{d}^{2})]}^{r}}{r-1} \nonumber \\
< &\frac{1-{d}^{1-r}}{r-1}. \label{25}
\end{align}
For each value of $r$, Eq.~(\ref{25}) yields an entanglement detection threshold value for $q$ which depends on $d$, and for a fixed value of the latter, decreases with increasing $r$.  This corresponds to improving detection performance with increasing $r$, a behavior which is known for $d$ equal to 2 and 3 \cite{GAM}.  Figure \ref{fig1} shows a plot of the threshold value $q$ versus the dimension $d$ for four different values of the order $r$, respectively increasing from top to bottom.  The improvement in entanglement detection for fixed $d$ and increasing $r$, as well as fixed $r$ and increasing $d$, is clearly in evidence in Fig.~\ref{fig1}.

The weakest performance obtains for $r=1$, top graph in Fig. 1, and can be found by taking the limit of Eq.~(\ref{25}) as $r \rightarrow 1$.  The result is $H[q+(1-q)/{d}^{2}])+ ({d}^{2}-1)H[q+(1-q)/{d}^{2}] < \ln(d)$, which corresponds to the choice of Shannon entropy function $H(\cdot)$ together with the set of generalized Bell states for the measurement of $d \otimes d$ Werner states \cite{ftn3}.    The best performance obtains for $r \rightarrow \infty$ in Eq.~(\ref{25}), which simplifies to $q > 1/(1+d)$, the known entanglement threshold for a the two-qudit Werner state.  This limiting case corresponds to the bottom graph in Fig.~\ref{fig1}.

The foregoing discussion of how well Corollary 1 does in detecting entanglement notwithstanding, the aim of this section is not to promote quasi-entropic detectors given in Corollaries 1-3 per se, but rather to emphasize the reach of the majorization results of \S III whence they are inherited.  In addition, the above example highlights the fact that, while the entanglement detection bounds given in Corollaries 1-3 are optimal for Schur-concave functions as a class, their performance need not be optimal for individual members of the class, a fact that was emphasized in paper I as well.

\section{Concluding Remarks}
In this paper we have developed entanglement detection criteria within the majorization formulation of uncertainty presented in paper I.  These are inherently stronger than similar scalar conditions in the sense that they are equivalent to and imply infinite classes of such scalar criteria.   Majorization relations in effect envelope the set of scalar measures of disorder based on Schur-concave functions, a huge set that includes the information-theoretically relevant class of quasi-entropic measures as a subset.   This enveloping property elucidates the exceptional effectiveness of majorization relations in dealing with problems of quantum information theory.

Entanglement detection criteria that can be experimentally implemented are especially useful in studies involving entangled microsystems, hence the importance of linear detectors and entanglement witnesses.   As already mentioned, Theorem 1 can be formulated as an entanglement witness inasmuch as it is a linear majorization condition on measurement results.  What may not be so obvious is that in principle the spectrum of a state is also experimentally accessible on the basis of the Lemma of \S IIIA.   That Lemma states that the spectrum of a quantum state is the supremum of probability vectors that may be deduced from the counter statistics of all possible rank 1 generalized measurements performed on the state.   In practice, this provides an experimental method for estimating the spectrum, since a high precision determination may require a search over a large number of possible measurements.

There is already an extensive literature on entanglement detection with many important results.  With the notable exception of the Nielsen-Kempe theorem, these results are not formulated within the majorization framework.  The present contribution in part serves to demonstrate the effectiveness of the majorization formulation of some of the existing methods.  One can reasonably expect a sharpening of the results of some of the other entanglement detection strategies when reformulated within the majorization framework.   It is also not unreasonable to expect useful, albeit computationally intractable, necessary and sufficient separability criteria to emerge from such formulations.

\begin{acknowledgments}
This work was in part supported by a grant from California State University, Sacramento.
\end{acknowledgments}

\appendix*

\section{majorization bounds are found on pure states}
Our task here is to establish that majorization uncertainty bounds can be reached on pure states.  We will prove this for two important cases, first for the uncertainty bound on all states of the system, and second on the class of separable states.   To avoid unnecessary clutter and technical distractions, we will limit the number of generalized measurements to three and the underlying Hilbert spaces to finite dimensions in the following treatment.

Suppose generalized measurements ($\mathrm{X}, \{ \hat{\mathrm{E}}_{\alpha}^{X} \}$), ($\mathrm{Y}, \{ \hat{\mathrm{E}}_{\beta}^{Y} \}$), and ($\mathrm{Z}, \{ \hat{\mathrm{E}}_{\beta}^{Z} \}$) are performed on a state $\hat{\rho}$ that is supported on a finite-dimensional Hilbert space.  What we will show below is that the search for the components of ${\mu}^{sup}$ which serve to define ${\mathscr{P}}_{sup}^{\mathrm{X} \oplus \mathrm{Y} \oplus \mathrm{Z}}$ can be limited to pure states.

By definition,
\begin{equation}
{\mathscr{P}}_{sup}^{\mathrm{X} \oplus \mathrm{Y} \oplus \mathrm{Z}}={\sup}_{\hat{\rho}}[{\mathscr{P}}^{\mathrm{X} \oplus \mathrm{Y} \oplus \mathrm{Z}}(\hat{\rho})], \label{A1}
\end{equation}
where
\begin{equation}
{\mathscr{P}}_{\alpha \beta \gamma}^{\mathrm{X} \oplus \mathrm{Y} \oplus \mathrm{Z}}(\hat{\rho})=\textrm{tr}({\hat{\mathrm{E}}}_{\alpha}^{X} \hat{\rho}) \textrm{tr}({\hat{\mathrm{E}}}_{\beta}^{Y} \hat{\rho}) \textrm{tr}({\hat{\mathrm{E}}}_{\gamma}^{Z} \hat{\rho}). \label{A2}
\end{equation}

We recall from \S IIA above and \S IIB, IVA and IVB of Paper I that the supremum defined by Eqs.~(\ref{A1}) and (\ref{A2}) is found by calculating ${\mu}^{sup}_{i}$ for $i=1,2,\ldots$ (${\mu}^{sup}_{0}=0$ by definition).   Furthermore, ${\mu}^{sup}_{1}$ is the maximum value of a single component of ${\mathscr{P}}^{\mathrm{X} \oplus \mathrm{Y} \oplus \mathrm{Z}}(\hat{\rho})$ as $\hat{\rho}$ is varied, ${\mu}^{sup}_{2}$ is the maximum value of the sum of two (different) components of ${\mathscr{P}}^{\mathrm{X} \oplus \mathrm{Y} \oplus \mathrm{Z}}(\hat{\rho})$ as $\hat{\rho}$ is varied, and ${\mu}^{sup}_{i}$ is the maximum value of the sum of $i$ (different) components of ${\mathscr{P}}^{\mathrm{X} \oplus \mathrm{Y} \oplus \mathrm{Z}}(\hat{\rho})$ as $\hat{\rho}$ is varied.  It is thus sufficient to prove that the said maximum for ${\mu}^{sup}_{i}$ is in fact realized on a pure state.

Let $\hat{\rho}={\sum}_{a}\, {\lambda}_{a}(\hat{\rho}) \mid {\psi}_{a} \rangle \langle {\psi}_{a} \mid$ be the principal ensemble representation for $\hat{\rho}$.  Then the above maximization process defines ${\mu}^{sup}_{i}$ as the maximum of
\begin{align}
{\sum}_{a,b,c}{\lambda}_{a}(\hat{\rho}){\lambda}_{b}(\hat{\rho}){\lambda}_{c}(\hat{\rho}) \big[ \langle {\psi}_{a} \mid {\hat{\mathrm{E}}}_{{\alpha}_{1}}^{X}\mid {\psi}_{a} \rangle \langle {\psi}_{b} \mid {\hat{\mathrm{E}}}_{{\beta}_{1}}^{Y}\mid {\psi}_{b} \rangle \nonumber \\ \times \langle {\psi}_{c} \mid {\hat{\mathrm{E}}}_{{\gamma}_{1}}^{Z}\mid {\psi}_{c} \rangle
+ \langle {\psi}_{a} \mid {\hat{\mathrm{E}}}_{{\alpha}_{2}}^{X}\mid {\psi}_{a} \rangle \langle {\psi}_{b} \mid {\hat{\mathrm{E}}}_{{\beta}_{2}}^{Y}\mid {\psi}_{b} \rangle \nonumber \\ \times \langle {\psi}_{c} \mid {\hat{\mathrm{E}}}_{{\gamma}_{2}}^{Z}\mid {\psi}_{c} \rangle
+ \ldots \,\,\,\,\,\,\,\,\,\,\,\,\,\,\,\,\,\,\,\,\,\,\,\,\,\,\,\,\,\,\,\,\,\,\,\,\,\,\,\,\,\,\,\,\,\,\,\,\,\,\,\,\,\,\,\,\,\,\,\,\,\,\,\,\,\,\,\, \nonumber \\
+ \langle {\psi}_{a} \mid {\hat{\mathrm{E}}}_{{\alpha}_{i}}^{X}\mid {\psi}_{a} \rangle \langle {\psi}_{b} \mid {\hat{\mathrm{E}}}_{{\beta}_{i}}^{Y}\mid {\psi}_{b} \rangle \langle {\psi}_{c} \mid {\hat{\mathrm{E}}}_{{\gamma}_{i}}^{Z}\mid {\psi}_{c} \rangle \big] \,\,\,\,\,\, \nonumber \\
+\,\, {\xi}_{i} \big(1-{\sum}_{a} {\lambda}^{\rho}_{a}\big)+{\sum}_{a,b} {\eta}_{i,ab} \big({\delta}_{ab}-\langle {\psi}_{b}\mid {\psi}_{a} \rangle\big),  \label{A3}
\end{align}
as the state vectors $\{\mid {\psi}_{a}\rangle \}$ which are the eigenstates of $\hat{\rho}$, the probabilities $\{ {\lambda}_{a}(\hat{\rho}) \}$ which constitute the spectrum of $\hat{\rho}$, and the Lagrange multipliers ${\xi}_{i}$ and ${\eta}_{i,ba}$ which serve to enforce the structure of ${\sum}_{a}\, {\lambda}^{\rho}_{a} \mid {\psi}_{a} \rangle \langle {\psi}_{a} \mid$ as an orthogonal ensemble are varied.  The symbol ${\delta}_{ab}$ in the above expression is the Kronecker delta, and the three sets of indices $\{ {({\alpha}, {\beta}, {\gamma})}_{1},  {({\alpha}, {\beta}, {\gamma})}_{2}, \dots, {({\alpha}, {\beta}, {\gamma})}_{i} \}$ are understood to be distinct, i.e., ${\alpha}_{1} \neq {\alpha}_{2} \neq \ldots \neq {\alpha}_{i}$, and similarly for the other two sets.  Note also that since $\langle {\psi}_{a}\mid {\psi}_{b} \rangle$ is in general a Hermitian matrix (in the indices $a$ and $b$), the Lagrange multipliers ${\eta}_{i,ba}$ may also be taken to constitute a Hermitian matrix.

A variation with respect to $\langle {\psi}_{a} \mid$ gives
\begin{equation}
{\lambda}_{a}(\hat{\rho}) \big[ {\hat{\mathcal{E}}}_{i}^{X}+ {\hat{\mathcal{E}}}_{i}^{Y}+{\hat{\mathcal{E}}}_{i}^{Z} \big]\mid {\psi}_{a} \rangle ={\sum}_{b} {\eta}_{i,ba} \mid {\psi}_{b} \rangle, \label{A4}
\end{equation}
where
\begin{equation}
{\hat{\mathcal{E}}}_{i}^{X}={\sum}_{k=1}^{i} \mathscr{P}^{\mathrm{Y}}_{{\beta}_{k}}(\hat{\rho})\mathscr{P}^{\mathrm{Z}}_{{\gamma}_{k}}(\hat{\rho})
{\hat{\mathrm{E}}}_{{\alpha}_{k}}^{X}, \label{A5}
\end{equation}
and ${\hat{\mathcal{E}}}_{i}^{Z}$ and ${\hat{\mathcal{E}}}_{i}^{Y}$ are defined similarly.  Note that the three operators introduced in Eqs.~(\ref{A4}) and (\ref{A5}) are Hermitian and positive.

Let ${\hat{\mathcal{E}}}_{i}={\hat{\mathcal{E}}}_{i}^{X}+ {\hat{\mathcal{E}}}_{i}^{Y}+{\hat{\mathcal{E}}}_{i}^{Z}$ and ${\mathcal{E}}_{i,aa}=\langle {\psi}_{a} \mid {\hat{\mathcal{E}}}_{i} \mid {\psi}_{a} \rangle$.  Then the maximum sought in (\ref{A3}), namely ${\mu}^{sup}_{i}$, is given by  ${\sum}_{a}{\lambda}_{a}(\hat{\rho}){\mathcal{E}}_{i,aa}/3=\textrm{tr}[{\hat{\mathcal{E}}}_{i}\hat{\rho}]/3$.  A key fact at this juncture is that the diagonal elements ${\mathcal{E}}_{i,aa}$ do not depend on $a$ and are all equal.  In other words, each pure state $\mid {\psi}_{a} \rangle$ in the ensemble contributes equally to ${\mu}^{sup}_{i}$.  This property follows from a variation of (\ref{A3}) with respect to ${\lambda}_{a}(\hat{\rho})$, which leads to
\begin{equation}
{\xi}_{i}=\langle {\psi}_{a} \mid {\hat{\mathcal{E}}}_{i}^{X}+ {\hat{\mathcal{E}}}_{i}^{Y}+{\hat{\mathcal{E}}}_{i}^{Z} \mid {\psi}_{a} \rangle={\mathcal{E}}_{i,aa}. \label{A6}
\end{equation}
Clearly then, each pure state $\mid {\psi}_{a} \rangle$ in the ensemble must realize the same maximum ${\mu}^{sup}_{i}$ as the entire ensemble $\hat{\rho}$.  We conclude therefore that the sought maximum can be found among the pure states of the system.

What if we are looking for ${\mu}^{sup}_{i}$ but with $\hat{\rho}$ limited to separable states?  It turns out that here too the search can be limited to pure states, which would be pure product states in this instance.   To establish this result, we will modify the foregoing analysis by stipulating that $\hat{\rho}={\sum}_{a}\, {q}_{a} \mid {\phi}^{A}_{a} \rangle \langle {\phi}^{A}_{a} \mid \otimes \mid {\phi}^{B}_{a} \rangle \langle {\phi}^{B}_{a} \mid$ representing a separable, bipartite state of parties $A$ and $B$ .   Then the corresponding ${\mu}^{sup}_{sep,i}$ is the maximum of
\begin{align}
{\sum}_{a,b,c}{q}_{a}{q}_{b}{q}_{c} \big[
\langle {\phi}^{B}_{a}\mid \otimes  \langle {\phi}^{A}_{a} \mid {\hat{\mathrm{E}}}_{{\alpha}_{1}}^{X}\mid {\phi}^{A}_{a} \rangle \otimes \mid {\phi}^{B}_{a} \rangle  \nonumber \\
\times \langle {\phi}^{B}_{b}\mid \otimes  \langle {\phi}^{A}_{b} \mid {\hat{\mathrm{E}}}_{{\beta}_{1}}^{Y}\mid {\phi}^{A}_{b} \rangle \otimes \mid {\phi}^{B}_{b} \rangle \nonumber \\
\times \langle {\phi}^{B}_{c}\mid \otimes  \langle {\phi}^{A}_{c} \mid {\hat{\mathrm{E}}}_{{\gamma}_{1}}^{Z}\mid {\phi}^{A}_{c} \rangle \otimes \mid {\phi}^{B}_{c} \rangle  \nonumber \\
+ \langle {\phi}^{B}_{a}\mid \otimes  \langle {\phi}^{A}_{a} \mid {\hat{\mathrm{E}}}_{{\alpha}_{2}}^{X}\mid {\phi}^{A}_{a} \rangle \otimes \mid {\phi}^{B}_{a} \rangle  \nonumber \\
\times \langle {\phi}^{B}_{b}\mid \otimes  \langle {\phi}^{A}_{b} \mid {\hat{\mathrm{E}}}_{{\beta}_{2}}^{Y}\mid {\phi}^{A}_{b} \rangle \otimes \mid {\phi}^{B}_{b} \rangle \nonumber \\
\times \langle {\phi}^{B}_{c}\mid \otimes  \langle {\phi}^{A}_{c} \mid {\hat{\mathrm{E}}}_{{\gamma}_{2}}^{Z}\mid {\phi}^{A}_{c} \rangle \otimes \mid {\phi}^{B}_{c} \rangle  \nonumber \\
+ \ldots \,\,\,\,\,\,\,\,\,\,\,\,\,\,\,\,\,\,\,\,\,\,\,\,\,\,\,\,\,\,\,\,\,\,\,\,\,\,\,\,\,\,\,\,\,\,\,\,\,\,\,\,\,\,\,\,\,\,\,\,\,\,\,\,\,\,  \,\,\,\, \nonumber \\
+ \langle {\phi}^{B}_{a}\mid \otimes  \langle {\phi}^{A}_{a} \mid {\hat{\mathrm{E}}}_{{\alpha}_{i}}^{X}\mid {\phi}^{A}_{a} \rangle \otimes \mid {\phi}^{B}_{a} \rangle  \nonumber \\
\times \langle {\phi}^{B}_{b}\mid \otimes  \langle {\phi}^{A}_{b} \mid {\hat{\mathrm{E}}}_{{\beta}_{i}}^{Y}\mid {\phi}^{A}_{b} \rangle \otimes \mid {\phi}^{B}_{b} \rangle \nonumber \\
\times \langle {\phi}^{B}_{c}\mid \otimes  \langle {\phi}^{A}_{c} \mid {\hat{\mathrm{E}}}_{{\gamma}_{i}}^{Z}\mid {\phi}^{A}_{c} \rangle \otimes \mid {\phi}^{B}_{c} \rangle  \nonumber \\
+ {\sum}_{a} {\eta}_{i,a}^{A} \big[ 1-\langle {\phi}^{A}_{a}\mid {\phi}^{A}_{a} \rangle ]
+{\sum}_{a} {\eta}_{i,a}^{B} \big[ 1-\langle {\phi}^{B}_{a}\mid {\phi}^{B}_{a} \rangle ]  \nonumber \\
+ {\xi}_{i} \big(1-{\sum}_{a} {q}_{a}\big), \,\,\,\,\,\,\,\,\,\,\,\,\,\,\,\,\,\,\,\,\,\,\,\,\,\,\,\,\,\,\,\,\,\,\,\,\,\,\,    \label{A7}
\end{align}
as the state vectors $ \{ \mid {\phi}^{A}_{a} \rangle \}$, $\{ \mid{\phi}^{B}_{a} \rangle \}$, the probabilities $\{ {q}_{a} \}$, and the Lagrange multipliers ${\eta}_{i,a}^{A}$, ${\eta}_{i,a}^{B}$, and ${\xi}_{i}$ which serve to enforce normalization conditions and probability conservation for the ensemble are varied.

Variations with respect to $ \langle {\phi}^{A}_{a} \mid$ and $ \langle {\phi}^{B}_{a} \mid$ now give
\begin{align}
{q}_{a} \big[ \langle {\phi}^{B}_{a}\mid ({\hat{\mathcal{E}}}_{i}^{X}+ {\hat{\mathcal{E}}}_{i}^{Y}+{\hat{\mathcal{E}}}_{i}^{Z}) \mid {\phi}^{B}_{a} \rangle \big]   \mid {\phi}^{A}_{a} \rangle &={\eta}_{i,a}^{A} \mid {\phi}^{A}_{a} \rangle  \nonumber \\
{q}_{a} \big[ \langle {\phi}^{A}_{a}\mid ({\hat{\mathcal{E}}}_{i}^{X}+ {\hat{\mathcal{E}}}_{i}^{Y}+{\hat{\mathcal{E}}}_{i}^{Z}) \mid {\phi}^{A}_{a} \rangle \big]   \mid {\phi}^{B}_{a} \rangle  &={\eta}_{i,a}^{B} \mid {\phi}^{B}_{a} \rangle ,  \label{A8}
\end{align}
where $({\hat{\mathcal{E}}}_{i}^{X},{\hat{\mathcal{E}}}_{i}^{Y},{\hat{\mathcal{E}}}_{i}^{Z})$ are defined as in Eq.~(\ref{A5}) et seq.

Equations (\ref{A8}) directly imply that ${\eta}_{i,a}^{A}={\eta}_{i,a}^{B}$, and as in the general case above, the maximum sought in (\ref{A7}), namely ${\mu}^{sup}_{sep,i}$, is found to equal  ${\sum}_{a}{q}_{a} {\mathcal{E}}_{i,aa}/3=\textrm{tr}[{\hat{\mathcal{E}}}_{i}\hat{\rho}]/3$, where ${\hat{\mathcal{E}}}_{i}={\hat{\mathcal{E}}}_{i}^{X}+ {\hat{\mathcal{E}}}_{i}^{Y}+{\hat{\mathcal{E}}}_{i}^{Z}$.   Furthermore, a variation with respect to the probabilities $\{ {q}_{a} \}$ shows the equality of the matrix elements ${\mathcal{E}}_{i,aa}$ just as in the general case above, whereupon we learn that each pure product state in the ensemble makes the same contribution to ${\mu}^{sup}_{sep,i}$.  In other words, the separable density matrix which maximizes (\ref{A7}) may be taken to be a pure product state \cite{noteext}.

It should be clear from the above arguments that the results hold for any number of generalized measurements as well as any number of parties in the multipartite state.   In summary, then, we have found that the joint majorization uncertainty bound, ${\mathscr{P}}_{sup}^{\mathrm{X} \oplus \mathrm{Y} \oplus \ldots \oplus \mathrm{Z}}$, for a set of generalized measurements performed on an arbitrary quantum state supported on a finite-dimensional Hilbert space can be found on the class of pure states of the system, and in the case of a separable states,  ${\mathscr{P}}_{sep;sup}^{\mathrm{X} \oplus \mathrm{Y} \oplus \ldots \oplus \mathrm{Z}}$ can be found on the class of pure product states of the system.

{}


\begin{thebibliography}{}
\bibitem{I}  M. Hossein Partovi, Phys. Rev. A \textbf{84}, 052117 (2011).   The literature on the subject can also be traced from the references given in this paper.
\bibitem{ftn0} Because the complete R\'{e}nyi family fails to be concave, we classified it as not belonging to the quasi-entropic class in Paper I.  Here we will use a more nuanced classification; see \S IV for further discussion, as well Ref. \cite{ftn2} below.
\bibitem{noteent}  Note that by an ``entangled state'' we mean any multipartite state that is not fully separable.
\bibitem{noteext} In technical language, these pure product states are extremal points in the (convex) set of separable states.
\bibitem{rem} The ensuing condition here is precisely equivalent to the separability condition in Eq.~(\ref{8}) of Nielsen and Kempe \cite{maj} which was obtained by a different method (to be discussed in \S IIIB).   Needless to say, this is not a surprising outcome since both methods are based on majorization relations.
\bibitem{ftn3}  Similar results have been found by other methods; see \cite{GIO}.
\end{thebibliography}
\end{document}